\documentclass[runningheads]{llncs}
\usepackage{booktabs}
\usepackage{cite}
\usepackage{url}
\usepackage{float}
\usepackage{xcolor}
\newcommand*{\tabindent}{\hspace{3mm}}
\begin{document}
\title{The Impact of Label Noise on a Music Tagger\thanks{This work is supported by the Austrian National Science Foundation (FWF P31988).}}

\author{Katharina Prinz\inst{}
\and
Arthur Flexer\inst{}
\and
Gerhard Widmer\inst{}
}
\authorrunning{K. Prinz, A. Flexer and G. Widmer}
%
\institute{Institute of Computational Perception, Johannes Kepler University Linz, Austria \\
\email{\{katharina.prinz,arthur.flexer,gerhard.widmer\}@jku.at}}
\maketitle              
\begin{abstract}
We explore how much can be learned from noisy labels in audio music tagging. Our experiments show that carefully annotated labels result in highest figures of merit, but even high amounts of noisy labels contain enough information for successful learning. Artificial corruption of curated data allows us to quantize this contribution of noisy labels.

\keywords{music tagging  \and label noise \and convolutional neural networks}
\end{abstract}
\section{Introduction}
\label{sec:introduction}
The necessity of annotated data for supervised learning often contrasts with the cost of obtaining reliable ground-truth in a manual fashion. Automated methods, on the other hand, simplify the annotation process and result in greater quantities of data with possibly noisy labels, which is why multiple approaches that investigate learning from noisy labels have been proposed~\cite{jiang2018curriculumfornoisy, natarjan2013learningnoisy, zhang2018celfornoisy, li2017noisydistillation, liu2015importancereweighting, sukhbaatar2015noisydnn}. In the DCASE2019 Challenge 
``Audio tagging with noisy labels and minimal supervision'', the question was raised whether noisy labels can also be useful when training a system to perform audio tagging \cite{fonseca2019dcase}. This work builds upon a submission to the DCASE2019 Challenge \cite{paischer2019dcasetechnical} and prior work \cite{2019lbd}, but with a focus particularly on audio samples with musical labels. We investigate the impact of training a music tagger solely on noisy labels, and find that even for a potentially high level of noise we clearly outperform a random baseline at a respectable performance level.

\section{Data}
\label{sec:data}
We use the musical subset of the data provided for Task 2 of the DCASE2019 Challenge \cite{fonseca2019dcase}. More precisely, out of the 80 classes in the original setup, our experiments are restricted to samples with one or several out of 12 different classes, containing musical instruments and male/female singing. In total, 3,967 training audio clips remain of which roughly 20\% (825) are curated, and 80\% (3,142) have noisy labels. In addition we reserve 712 audio clips with curated labels for testing. \textit{Curated} audio clips have been carefully manually annotated \cite{fonseca2017freesound}, whereas \textit{noisy} labels are the result of automated heuristics \cite{fonseca2019dcase}. As annotations are only available at clip-level without time stamps, we speak of weak labels. The average number of labels per curated training clip is 1.03, for curated test clips it is 1.06 and for noisy clips 1.15.

\section{Methods}
\label{sec:methods}
We use a Convolutional Neural Network (CNN) with eight convolutional and three average-pooling layers as proposed in \cite{paischer2019dcasetechnical}. The inputs of the CNN are 96-bin Mel-spectrograms, transformed to decibel scale. For computing the spectrograms, we use a Fast-Fourier Transform (FFT) size of 2048, a hop-size of 512 and a minimum frequency of 40 Hertz (Hz). Prior to feature computation, raw audio is resampled to 16 kHz. Due to improved results in preliminary experiments, spectrograms are not normalised before being fed into the CNN. In the learning procedure, we use an Adam optimizer with an initial learning rate of 0.001. After 80 epochs, the learning rate is reduced by a factor of 10, and training is continued for 20 more epochs. Batch normalisation and drop-out are used against overfitting. For training, we use one random 3 second snippet per clip; for testing, on the other hand, the full audio is used. In case the raw audio is too short, i.e. shorter than 3 seconds for training samples, circular padding is used. 

\section{Results}
\label{sec:results}
Task 2 of the DCASE2019 Challenge was to maximise the performance of an audio tagging system on weakly multi-labelled data by exploiting both curated and noisy labels. In contrast to this, we focus on the impact these two types of labels have individually. Particularly, we take a closer look at the contribution of noisy labels in the task of music tagging. Before evaluating the performance of our system on different training data, we tune hyper-parameters (cf. section \ref{sec:methods}) on a separate curated validation set. After this, we train several models; first we use training data with curated labels only, and secondly with noisy labels only. Furthermore, we compare their performance with a random baseline and a variation of the model with intentionally corrupted labels.

To measure the performance of the tagging system, we use the mean average precision (MAP) and the mean area under ROC curves (MAUC) (cf. \cite{shah2018closerlook}). We repeat our experiments 5 times for different random CNN initializations, and show mean and standard deviation of MAP and MAUC on the test set over all runs in Table \ref{tab:results}. Additionally, we perform paired sample t-tests to determine the significance of different performances. In what follows, differences are denoted as \textit{significant} whenever $\left| t \right| > t_{(99,df=4)} = 4.604$.

\begin{table}
\begin{center}
\begin{tabular}{l c c}
\toprule
\textbf{Training Data} & \tabindent\textbf{MAP} & \tabindent\textbf{MAUC} \\
\midrule
Curated Labels & \tabindent0.767 $\pm$ 0.005 & \tabindent0.913 $\pm$ 0.004 \\
Noisy Labels & \tabindent0.665 $\pm$ 0.013 & \tabindent0.846 $\pm$ 0.010 \\
\midrule
Random Labels & \tabindent0.265 $\pm$ 0.006 & \tabindent0.502 $\pm$ 0.007 \\
Corrupted Labels (r=70 \%) & \tabindent0.638 $\pm$ 0.019 & \tabindent0.852 $\pm$ 0.009 \\
\bottomrule
\end{tabular}
\vspace{3mm}
\caption{Mean $\pm$ std.~deviation over 5 runs of models with different training data.}
\vspace{-1cm} 
\label{tab:results}
\end{center}
\end{table}

The first two lines in Table \ref{tab:results} show the performance of our CNN trained on either data with curated labels only, or noisy labels only. Lines 3 and 4 relate to a random baseline and a corrupted version of the curated training set. 

More precisely, for line 3 and 4 in Table \ref{tab:results} we train the model on curated data once more, but with two different modifications to the curated labels. First, we create a \textit{random} baseline with an intact label distribution by shuffling the labels of the training data. Furthermore, we estimate the unknown level of noise present in the noisy dataset by performing the second modification in line 4, for which we intentionally corrupt a certain percentage of curated labels until we reach a similar performance as in line 2. This is done by replacing one single random tag by a random but differing new tag for $r\%$ of curated training data, regardless of the original number of tags for a particular clip. In other words, all clips remain with the same number of tags as before, but with exactly one wrong tag. Note here that due to the low average number of tags per clip (cf. section \ref{sec:data}), the resulting level of noise will be closely related to the $r$ value.

Training our CNN with curated audio clips results in a mean performance of 0.767 MAP and 0.913 MAUC as shown in Table \ref{tab:results}. This is, for both metrics, a significant difference to the random baseline and to our model trained on data with noisy labels only. Similarly, the difference between the model trained solely on noisy labels (with an average MAP of 0.665 and MAUC of 0.846) and our baseline is statistically significant. If we decrease the number of training samples with noisy labels to correspond to the lower amount of 825 curated training samples, the MAP and MAUC decrease to $0.587 \pm 0.012$ and $0.795 \pm 0.016$, respectively (not shown in Table \ref{tab:results}). This benefit of using large quantities of data with noisy labels is in line with previous results on comparable data \cite{fonseca2019learningsound}.

For the last line in Table \ref{tab:results}, we show the result of training on data with a noise level of $r=70\%$; this comes close to the performance of training on the actual noisy dataset (not a statistically significant difference). Starting with $r=0\%$ and increasing this factor with a step-size of $5\%$, the MAP and MAUC show the first significant decrease when reaching a level of $50\%$ noise. At this point, the average MAP (MAUC) is reduced from 0.767 (0.913) to 0.736 (0.903) (not shown in Table \ref{tab:results}). Training the CNN on corrupted labels only, i.e. $r=100\%$, decreases the two metrics to on average 0.206 and 0.388 respectively, which for the MAUC is a significant difference compared to the random baseline.

\section{Discussion and Conclusion}
\label{sec:conclusion}
In conclusion, we see that even though training a music tagger on a set of curated audio samples leads to the best performance, a model trained on very noisy labels still outperforms a random baseline significantly, with figures of merit actually much closer to the carefully curated scenario. This is particularly interesting as noisy and curated clips have the same set of classes, but originate from different sources \cite{fonseca2019dcase}. Being based on Freesound (\url{www.freesound.org}) content, curated audio files often contain isolated sound samples of a class, while noisy files tend to be of a more composite nature, as they consist of Flickr video soundtracks.

To explore the unknown level of noise in the noisy scenario provided for the DCASE2019 challenge \cite{fonseca2019dcase}, we performed additional experiments in which we trained on curated data with intentionally corrupted labels of a certain percentage (cf. \cite{shah2018closerlook}). Introducing this controlled amount of noise suggests that the noisy dataset we tried to learn from possibly contains a relatively high amount of $70\%$ wrong labels, although we do not yet have an approximation of how the domain mismatch influences the differences in performance of curated and noisy training data. Nevertheless we were able to show that a  weak multi-label audio-tagger trained solely on noisy labels can not only perform significantly better than a random baseline but at a respectable performance level, even in the case of a domain mismatch and a potential high level of noise.

%
%
%

\begin{thebibliography}{8}


\bibitem{fonseca2019dcase}
Fonseca, E., Plakal, M., Font, F., Ellis, D.P.W. and Serra, X.: Audio tagging with noisy labels and minimal supervision. In: Proc.\ of the Detection and Classification of Acoustic Scenes and Events 2019 Workshop, pp. 69--73. New York University (2019).

\bibitem{fonseca2019learningsound}
Fonseca, E., Plakal, M., Ellis, D.P.W., Font, F., Favory, X., Serra, X.: Learning sound event classifiers from web audio with noisy labels. In: Proc.\ of the IEEE Intern.\ Conf.\ on Acoustics, Speech and Signal Processing, pp. 21--25. IEEE (2019). 

\bibitem{fonseca2017freesound}
Fonseca, E., Pons, J., Favory, X., Font, F., Bogdanov, D., Ferraro, A., Oramas, S., Porter, A., Serra, X.: Freesound datasets: a platform for the creation of open audio datasets. In: Proc.\ of the 18th Intern.\ Society for Music Information Retrieval, pp. 486--493. ISMIR (2017).

\bibitem{jiang2018curriculumfornoisy}
Jiang, L., Zhou, Z., Leung, T., Li, L.J., Fei-Fei, L.M.: MentorNet: Learning data-driven curriculum for very deep neural networks on corrupted labels. In: Proc.\ of the 35th Intern.\ Conf.\ on Machine Learning, pp. 2309--2318. PMLR (2018). 

\bibitem{li2017noisydistillation}
Li, Y., Yang, J., Song, Y., Cao, L., Luo, J. and Li, L.J.: Learning from noisy labels with distillation. In: Proc.\ of the IEEE Intern.\ Conf.\ on Computer Vision, pp. 1910--1918. IEEE (2017).

\bibitem{liu2015importancereweighting}
Liu, T., Tao, D.: Classification with noisy labels by importance reweighting. IEEE Transactions on pattern analysis and machine intelligence \textbf{38}(3), 447--461 (2015).

\bibitem{natarjan2013learningnoisy}
Natarajan, N., Dhillon, I.S., Ravikumar, P.K., Tewari, A.: Learning with noisy labels. In: Advances in neural information processing systems, pp. 1196--1204. Curran Associates (2013).

\bibitem{paischer2019dcasetechnical}
Paischer, F, Prinz, K., Widmer, G.: Audio tagging with convolutional neural networks trained with noisy data. DCASE2019 Challenge (2019).

\bibitem{2019lbd}
Prinz, K., Flexer, A.: Weak multi-label audio-tagging with class noise. Late-Breaking/Demo ISMIR (2019).

\bibitem{shah2018closerlook}
Shah, A., Kumar, A., Hauptmann, A.G., Raj, B.: A closer look at weak label learning for audio events. CoRR \textbf{abs/1804.09288} (2018).

\bibitem{sukhbaatar2015noisydnn}
Sukhbaatar, S., Fergus, R.: Learning from noisy labels with deep neural networks. In: Proc.\ of the 3rd Intern.\ Conf.\ on Learning Representations (2015).

\bibitem{zhang2018celfornoisy}
Zhang, Z., Sabuncu, M.: Generalized cross entropy loss for training deep neural networks with noisy labels. In: Advances in neural information processing systems, pp. 8778--8788. Curran Associates (2018).

\end{thebibliography}
%

\end{document}